\documentclass[twocolumn,superscriptaddress,nofootinbib,floatfix,aps,prb,citeautoscript,longbibliography]{revtex4-2}
\usepackage[utf8]{inputenc}
\usepackage[T1]{fontenc}
\usepackage{lineno}

\usepackage{graphicx}
\usepackage{color}
\usepackage{array}
\usepackage{dcolumn}
\usepackage{bm}
\usepackage{multirow}
\usepackage{chemformula}
\let\ce\ch
\usepackage{tabularx}
\usepackage{comment}
\usepackage{tikz}
\usepackage{xstring}

\usepackage[normalem]{ulem}
\usepackage{amsmath}
\usepackage{float}
\usepackage{booktabs}
\usepackage{tabularray}
\usepackage{orcidlink}
\usepackage{cleveref}
\usepackage{subcaption}

\newcommand{\Tc}{\ensuremath{T_\textrm{c}}}

\newcommand{\epsinf}{\ensuremath{\epsilon_{\infty}}}

\newcommand{\add}[1]{#1%
}
\newcommand{\remove}[1]{\ignorespaces%
}

\begin{document}

\newcommand{\bochum}{Research Center Future Energy Materials and Systems of the University Alliance Ruhr and Interdisciplinary Centre for Advanced Materials Simulation, Ruhr University Bochum, Universit\"{a}tsstra{\ss}e 150, D-44801 Bochum, Germany}

\author{Paulo R. Pires\orcidlink{0009-0005-7598-4862}}
\affiliation{\bochum}
\author{Pierre-Paul De Breuck\orcidlink{0000-0002-3173-2058}}
\affiliation{\bochum}
\author{Mauro Fava\orcidlink{0000-0002-3461-1585}}
\affiliation{\bochum}
\author{Hai-Chen Wang\orcidlink{0000-0002-2892-5879}}
\affiliation{\bochum}
\author{Miguel A. L. Marques\orcidlink{0000-0003-0170-8222}}
\email{miguel.marques@rub.de}
\affiliation{\bochum}
\date{\today}

\title{Encoding Orbital-Projected Densities of States in Graph Neural Networks for Materials Property Prediction}

\begin{abstract}
{\centering\textbf{Abstract}\par}\vspace{0.5\baselineskip}

Graph neural networks have become the dominant machine-learning architecture for predicting materials properties from crystal structures. Yet the initialization of atomic node features has received comparatively little attention, and conventional approaches rely on static elemental descriptors that carry no information about the quantum-mechanical electronic environment of each atom in its crystalline host. Here we show that augmenting atomic node representations with site-projected orbital density of states (pDOS) fingerprints, computed directly from density functional theory calculations, yields systematic and substantial improvements in predictive performance. These representations are fused with Pettifor elemental embeddings at each atomic site before message passing. For the superconducting critical temperature \Tc\ and the optical dielectric constant \epsinf, the pDOS augmentation reduces prediction errors by 22.8\% and 27.9\%, respectively, relative to the elemental-descriptor baseline. These improvements are comparable to those achieved by doubling the training-set size. The gains are, however, contingent on training-set size. For the magnetic exchange energies of Heusler compounds, a substantially smaller dataset, the improvement is reduced, indicating that pDOS augmentation is most effective when the training data exceeds the length of the pDOS feature vector. We introduce an interpretable spectral attention-gating mechanism that reveals that the model autonomously learns to prioritize the orbital channels and energy windows most physically relevant to each target property. These results establish pDOS-augmented graph nodes as a broadly applicable strategy for infusing first-principles electronic-structure knowledge into graph networks, opening a practical route to high-accuracy property prediction in data-scarce regimes.
\end{abstract}

\maketitle
\twocolumngrid

\section{Introduction}
\label{sec:intro}

In recent years, machine-learning models trained on first-principles databases have emerged as a practical solution to the computational bottleneck posed by density functional theory (DFT)~\cite{Hohenberg1964,Kohn1965}, offering property predictions at a fraction of the cost while retaining much of the accuracy of the underlying quantum-mechanical calculations~\cite{Butler2018,Schmidt2019,Jha2018}. Such data-driven approaches benefit from ever-growing materials databases. However, database sizes grow slowly relative to the demand for predictions on new functional materials, motivating the development of more data-efficient representations and architectures.

Among all data-driven approaches for materials discovery, graph neural networks (GNNs) have become the dominant architecture~\cite{Jain2024,Xie2018}. Message-passing operations aggregate neighborhood information iteratively, allowing the network to learn representations of the local chemical environment directly from the crystal graph without requiring hand-crafted descriptors~\cite{Xie2018}. Subsequent work has extended this framework in two directions. First, three-body angular interactions have been incorporated either through directional message passing on pairs of bonds~\cite{Gasteiger2020} or through line-graph formalisms in which bond pairs sharing an atom form the edges of a secondary graph~\cite{Choudhary2021}. Second, equivariant architectures have enforced physical symmetries explicitly, yielding substantial accuracy gains for forces and energy surfaces~\cite{Batzner2022,Batatia2022,Schutt2021}. Together, these advances have dramatically improved accuracy for properties sensitive to local coordination geometry.

Despite these architectural advances, the representation of atomic node features has received comparatively little attention. In most crystal GNNs, nodes are initialized from elemental look-up-table descriptors (such as the atomic number, electronegativity, covalent radius, valence electron count) or learned elemental representations derived from physical properties~\cite{Xie2018,Chen2019}. Such embeddings carry no information about the electronic state of a particular atom \emph{within a given crystal}, as they are identical for the same chemical species regardless of whether it sits in a metallic, semiconducting, or magnetic host, or whether it is bonded to oxygen, to nitrogen, or to a transition metal. Some works have approached this problem by implementing dynamic embeddings on the graph nodes~\cite{Gupta2026}. \add{Although these embeddings adapt to the local environment, they are still learned from the geometry alone. Electronic-structure information has been used as a direct input in only a few cases. Fung \textit{et al.} introduced DOSnet~\cite{Fung2021Jan}, in which convolutional neural networks extract features from the site- and orbital-projected densities of states of the surface and adsorbate atoms to predict adsorption energies, without a graph representation of the structure. More recently, Karimitari \textit{et al.}~\cite{Karimitari2025Dec} incorporated electronic-structure information into the nodes of the ALIGNN graph~\cite{Choudhary2021}, with each node initialized with the PBE eigenvalue of a single Kohn--Sham state, together with the fractions of that state projected onto the $s$, $p$, and $d$ orbitals of the corresponding site, in order to predict the same eigenvalue at the hybrid-functional level. These works demonstrate the value of electronic-structure inputs, but they are tailored to specific architectures and targets, namely adsorption energies with convolutional networks or single-particle eigenvalue corrections. To the best of our knowledge,} \remove{However,} no existing approach \add{encodes the energy-resolved electronic structure} \remove{directly encodes the actual quantum-mechanical electronic state} of each atomic site \add{into the node features of a crystal graph for general property prediction}. \add{This is} information shaped by hybridization, charge transfer, and crystal-field effects, which \add{conventional GNNs} must therefore \add{infer} \remove{be inferred} implicitly from structural data alone.

The site and orbital-projected density of states (pDOS) provides a natural and physical route to encoding this missing information. By decomposing the total electronic density of states into orbital-resolved contributions at each atomic site, the pDOS captures a fingerprint computed directly from quantum-mechanical first-principles calculations. Importantly, the pDOS is both a site-specific and compound-specific quantity, as it changes with composition, structure, oxidation state, and bonding environment, information that no static elemental descriptor can reproduce. Furthermore, and in contrast to the band-structure, it is a real space quantity that can be naturally integrated in a GNN.

In the present work, we investigate how encoding this information into graph nodes can augment the accuracy of property inference. We validate our approach using \textsc{tuga-sp}, a new invariant crystal graph neural network (GNN) architecture employing dual-view representations: the primal graph and line graph, with message-passing over the line graph to capture angular information. We demonstrate that incorporating pDOS node features improves predictive performance and that the benefit becomes especially pronounced at smaller dataset sizes. \add{We further show that the improvement is not specific to our architecture, as adding the same pDOS embedding to ALIGNN~\cite{Choudhary2021, Lee2026Sep} and to BEE-NET~\cite{Gibson2026Jan}  we obtain comparable gains. The improvement also extends to electron--phonon, thermodynamic, and, to a lesser degree, elastic properties.} To better interpret how the model perceives this feature for the inference of properties, we implement attention maps over the pDOS encoding that reveal the spectral features the model prioritizes, offering a window into non-trivial physical relationships captured by the model.

\section{Results}
\label{sec:results}

\subsection{Superconducting \Tc\ and dielectric constants}

\begin{figure}[tbh]
    {\centering
    \includegraphics[width=0.5\textwidth]{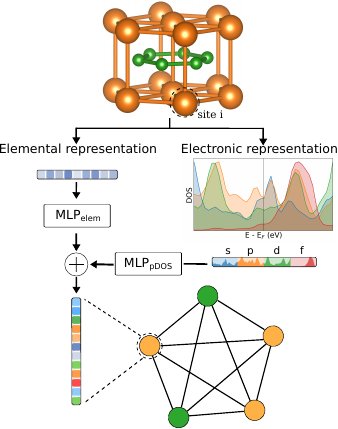}}
    \caption{Schematic overview of the pDOS node-encoding strategy. For each atomic site the orbital-projected density of states ($s$, $p$, $d$, $f$ channels) is computed from DFT and summed with the Pettifor elemental embedding before being passed to the \textsc{tuga-sp} message-passing network.}
    \label{img:method_schem}
\end{figure}

In order to implement our approach,\add{ summarized in \cref{img:method_schem}}, we retrieve the site-resolved orbital-projected density of states (pDOS) from VASP~\cite{Kresse1996, Kresse1996_2} calculations present in the Alexandria database~\cite{Cavignac2025}. These are then interpolated onto a uniform energy grid centered on the Fermi energy, producing a fixed-length fingerprint for each atomic site. Orbital channels absent for a given species are zero-padded, ensuring a consistent feature dimensionality across all elements. This pDOS fingerprint is then projected into a latent space and summed with an encoded static elemental representation, also projected into the same dimension. Throughout this work we use the Pettifor elemental embeddings~\cite{Cerqueira2026}, a non-orthogonal species representation based on chemical similarity. The resulting augmented node features therefore represent both the intrinsic identity of the atomic species and its site-specific electronic structure within the compound.

We employ \textsc{tuga-sp} \add{(see Methods)} to predict two target properties: the theoretical superconducting critical temperature (\Tc) \add{and} the optical dielectric constant (\epsinf)\remove{, and the nearest-neighbor isotropic exchange interaction ($J$) of the Heusler compounds}. The \epsinf\ presents a particular challenge for regression models, as its distribution spans several orders of magnitude and diverges for metallic systems. To circumvent this, the model is trained to predict the inverse dielectric constant \epsinf$^{-1}$, which vanishes naturally in the metallic limit. \add{Other properties will be used later to evaluate the generalization of the implementation of the pDOS.} \remove{The exchange energies will be discussed later in Sec.\ref{sec:heusler}.}

While conventional superconductivity is inherently associated with metallic systems, the dielectric constant is typically studied in insulating and semiconducting ones. The absence/presence of an electronic gap defines two regimes requiring different treatments of the pDOS encoding. For \Tc\, the relevant electronic structure is concentrated near the Fermi level, where Cooper pairing originates. Accordingly, the pDOS is binned onto a grid spanning $[-1, +1]$~eV relative to the Fermi energy $E_F$, with a uniform bin width of 0.1~eV, yielding a 21-point fingerprint per orbital channel per site. For \epsinf, the pDOS is interpolated onto two energy windows, in the valence-band $[\varepsilon_\mathrm{VBM} - 1,\, \varepsilon_\mathrm{VBM}]$~eV and in the conduction-band $[\varepsilon_\mathrm{CBM},\, \varepsilon_\mathrm{CBM} + 1]$~eV, where $\varepsilon_\mathrm{VBM}$ and $\varepsilon_\mathrm{CBM}$ denote the valence band maximum and conduction band minimum, respectively. The indirect band gap $\Delta = \varepsilon_\mathrm{CBM} - \varepsilon_\mathrm{VBM}$ is additionally concatenated as a scalar feature, providing the model with explicit information about the gap magnitude, which is strongly correlated with the dielectric response~\cite{Penn1962, Marques2011,trinquetOpticalMaterialsDiscovery2024}.

Hyperparameter optimization was performed independently for both target properties, considering two node-embedding variants, one augmented with the pDOS and one relying solely on the Pettifor elemental representation. This yielded four optimization studies, one per property per embedding type, each comprising 200 trials using the Optuna hyperparameter framework~\cite{Optuna2019}. The search minimized the mean absolute error (MAE) for \Tc\ and the mean squared error (MSE) for \epsinf$^{-1}$. The \Tc\ dataset comprises in total 34\,880 entries and the \epsinf\ dataset contains 20\,000 entries, the data was \add{randomly} divided into a 8:1:1 ratio for training, validation and test respectively.

Incorporating the pDOS node features yields a consistent improvement in predictive performance across both properties. For \Tc, the test MAE decreases by 22.8\% relative to the Pettifor-only baseline, going from 1.27~K to 0.98~K, while for \epsinf$^{-1}$ the corresponding reduction is 27.9\%, from 2.40$\times 10^{-4}$ to 1.73$\times 10^{-4}$ \add{of mean squared error}. This gain substantially exceeds what is typically observed when substituting the Pettifor embedding for another static elemental representation~\cite{Cerqueira2026}, showing that the benefit does not arise merely from a richer elemental encoding but from explicitly informing the model about the electronic states of each atomic site within its crystalline environment, which is invisible to any elemental representation.

\begin{figure*}[tbh]
    \centering
    \includegraphics[width=\textwidth]{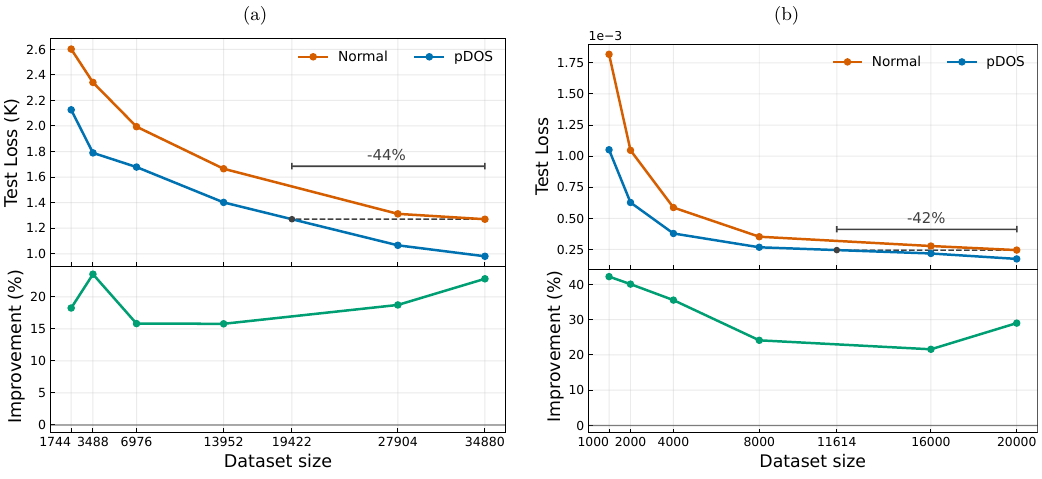}
    \caption{Error in the test set as a function of training set size for the pDOS-augmented model and the elemental-descriptor baseline, along with the relative improvement gained by incorporating orbital projections. (a)~Conventional superconducting critical temperature \Tc; (b)~Inverse dielectric constant (\epsinf$^{-1}$). The dashed lines indicate the baseline error at full dataset size and the corresponding interpolated training-set size at which the pDOS model achieves equivalent performance.}
    \label{img:dataset_difference}
\end{figure*}

To assess how the relative benefit of the pDOS embedding scales with dataset size, we grew the training set progressively, starting from 5\% of the available data and doubling the subset at each step while retaining all previously included materials. The same optimized hyperparameters were applied uniformly across all dataset sizes. The results depicted in \cref{img:dataset_difference} reveal that the performance gain from the pDOS embedding is especially important in the low-data regime. For \Tc, the pDOS model trained on $\sim$7\,000 entries (20\% of the full set) achieves the same prediction error as the baseline model trained with the $\sim$14\,000 entries (40\%), demonstrating that quantum-mechanically informed node initialization can effectively substitute for a factor-of-two increase in the training data. Evaluated at full dataset size, adopting pDOS features is equivalent to having approximately $1.8\times$ as much training data for \Tc.
The advantage is even more pronounced for the dielectric constant, where the relative gain exceeds 35\% for the smallest subsets and remains above 20\% across all tested dataset sizes. When considering the full dataset in the baseline scenario, using the pDOS feature is equivalent of using $1.7\times$ more data points to achieve the same error. \add{We stress again that it is known that the dielectric constant is highly correlated with the magnitude of the band gap~\cite{Marques2011} (see also the section ``Attention over orbital projections''). Given this, we trained a new model, with the same dataset and the same hyperparameters as the pDOS model, in which the only site property encoded alongside the Pettifor representation is the band gap itself, i.e.\ the last entry of the pDOS feature vector, replicated over all sites. The results are collected in \cref{tab:gap_only}. The band-gap-only model recovers roughly one third of the pDOS improvement in MAE (and about half in MSE), while the remainder requires the full orbital-resolved projections. The band gap is therefore a substantial, but not sufficient, part of the information that the pDOS embedding supplies for this property.}

\begin{table}[htb]
    \add{
    \caption{\add{The loss error and gain in accuracy for the dieletric constant when considering only the input of band gap and band gap with pdos, compared to the baseline scenario of no eletronic information.}}
    \label{tab:gap_only}
    \begin{tabular}{lccc}
    \toprule 
    Site encoding & MAE & MSE  & Gain (MSE) \\
    \midrule
    Baseline             & 6.94$\times 10^{-3}$ & 2.40$\times 10^{-4}$  & --      \\
    Band gap only        & 6.60$\times 10^{-3}$ & 2.11$\times 10^{-4}$  & -12.1\%   \\
    Full pDOS            & 5.94$\times 10^{-3}$ & 1.73$\times 10^{-4}$  & \textbf{-27.9\%}  \\
    \bottomrule
    \end{tabular}
    }
\end{table}

In both plots we see that as the dataset increases, the improvement in accuracy first decreases, and finally increases again. We believe that this is ultimately a consequence of using the optimized hyperparameters of the full dataset. Beyond that we also note that, as the training set size decreases below $\sim$500 entries, a value comparable in size to the pDOS feature vector itself, the model can no longer extract reliable statistical patterns from the orbital projections, and the pDOS and baseline variants yield similar errors. We additionally examined the sensitivity of model performance to the width of the pDOS energy window. Accuracy barely increases with wider windows, though the optimal range might be property-dependent.

\add{To understand how other models behave under the implementation of pDOS, we trained ALIGNN~\cite{Choudhary2021, Lee2026Sep} and BEE-NET~\cite{Gibson2026Jan} (crystal structure only) using the exact same data as \textsc{tuga-sp}. We used the same compounds and the same train/validation/test partitions, with the L1 loss for \Tc\ and the MSE loss for the dielectric constant. The pDOS enters these architectures exactly as it does in our model leaving every other component of the original implementations untouched. For each architecture we compare four node initializations: the atomic embedding of the original model with and without the pDOS, and the Pettifor representation with and without the pDOS. Given the high cost of performing hyperparameter optimization for all these models, we decided to use the default hyperparameters of each original implementation.}

\begin{table*}[htb]
\add{
\centering
\caption{Effect of the pDOS embedding on other architectures, trained on the same data and partitions as our model and with the default hyperparameters of each original implementation. The error is the MAE for \Tc\ and the MSE for the dielectric constant and the gain is its reduction with respect to the same architecture without the pDOS.}
\label{tab:other_models}
\begin{tabular}{llcccc}
\toprule
& & \multicolumn{2}{c}{\Tc} & \multicolumn{2}{c}{Dielectric constant} \\
\cmidrule(lr){3-4}\cmidrule(lr){5-6}
Model & Node initialization & MAE (K) & Gain & MSE ($10^{-4}$) & Gain (MSE) \\
\midrule
\multirow{4}{*}{ALIGNN}
 & CGCNN (original)   & 1.27  & --       & 3.40 & --       \\
 & CGCNN + pDOS       & 0.958 & -24.3\%   & 2.23 & -34.3\%   \\
 & Pettifor           & 1.20  & --       & 3.12 & --       \\
 & Pettifor + pDOS    & \textbf{0.940} & -21.9\% & 2.27 & -27.1\% \\
\midrule
\multirow{4}{*}{BEE-NET}
 & one-hot (original) & 1.58  & --       & 9.32 & --       \\
 & one-hot + pDOS     & 1.25  & -21.0\%   & 6.92 & -25.7\%   \\
 & Pettifor           & 1.51  & --       & 9.85 & --       \\
 & Pettifor + pDOS    & 1.20  & -20.6\%   & 6.97 & -29.3\%   \\
\midrule
This work & Pettifor + pDOS & 0.98 & -22.8\% & \textbf{1.73} & -27.9\% \\
\bottomrule
\end{tabular}
}
\end{table*}

\add{The results, summarized in \cref{tab:other_models}, show that the benefit of the pDOS embedding is not specific to our architecture. The improvement is systematic across models, properties and node initializations, and amounts to a reduction of the loss between 21\% and 24\% for \Tc\ and between 26\% and 34\% for the dielectric constant. The choice of the elemental representation, on the contrary, is of secondary importance, as replacing the original atomic embedding by the Pettifor representation changes the error by a few percent at most, and the ordering of the two is not even consistent between properties. For \Tc\ the gain obtained from the pDOS is moreover larger than the spread between the architectures themselves. BEE-NET with the pDOS is as accurate as ALIGNN without it, and ALIGNN with the pDOS surpasses marginally our own model trained on the same data. For the dielectric constant the situation is different, since the gap between the two architectures is considerably larger than the gain provided by the pDOS, and the ordering of the models is therefore unchanged.}

\add{BEE-NET was originally developed to predict the Eliashberg spectral function. For a better comparison, we summed the pDOS embedding to its mass-weighted one-hot node embedding. The results are shown in \cref{tab:bee-net}. The error decrease obtained by introducing the pDOS (18.3\%) is comparable to that obtained by introducing the phonon DOS (18.8\%), and combining both yields the best result, with a decrease of 24.7\%. The absolute errors differ from those reported in the original BEE-NET work, which is expected given the differences between the datasets. The dataset used for BEE-NET consists mostly of compounds with low \Tc, whereas ours includes, for instance, a larger number of higher \Tc\ hydrides. The phonon spectra of these compounds extend to much higher frequencies, so $\alpha^2F(\omega)$ had to be predicted on a frequency grid with considerably more points than in the original work, and no hyperparameter optimization was performed for this setting. Nonetheless, all four models were trained and tested under the same conditions.}

\begin{table}[htb]
\add{
\caption{\add{The MAE for \Tc{} and gain in accuracy for BEE-NET considering the crystal structure only model (CSO) and the Coarse Phonon Density of states model (CPD), with and without the pDOS embedding. The usage of the phonon DOS achieves a gain in performance comparable to the embedding of the pDOS, being the best case when both descriptors are used.}}
\label{tab:bee-net}
\begin{tabular}{lccc}
\toprule 
Model & MAE (K) & Gain \\
\midrule
CSO (baseline)   & 1.86 & - \\
CSO + pDOS      & 1.52  & -18.3\% \\
CPD           & 1.51 & -18.8\% \\
CPD + pDOS  & \textbf{1.40} & \textbf{-24.7\%} \\
\bottomrule
\end{tabular}
}
\end{table}

\subsection{Other properties}

\add{We finally evaluate the implementation of the pDOS for other properties to assess its generality. For this we predicted a group of properties with and without the pDOS: the logarithmic average phonon frequency $\omega_{\log}$ and the electron-phonon coupling constant $\lambda$ from the same \Tc\ dataset, the formation energy and the energy above the convex hull of thermodynamic stability from 50\,000 randomly selected compounds in the Alexandria database~\cite{Cavignac2025}, and the shear, bulk and Young moduli from the benchmark dataset used in~\cite{MatCalc}. The last set is the most demanding test of the idea, since the elastic response is a ground-state property of the total energy surface and is not expected to be governed by the states close to the Fermi level that the pDOS window resolves. The results are collected in \cref{tab:properties}.}

\begin{table*}[htb]
\centering
\add{
\caption{Test error with and without the pDOS embedding for a set of additional properties. Each pair of models shares the dataset, the partitions and the training protocol of the corresponding models. The gain is the reduction of the MAE brought by the pDOS.}
\label{tab:properties}
\begin{tabular}{lccccc}
\toprule
& \multicolumn{2}{c}{Baseline} & \multicolumn{2}{c}{With pDOS} & \\
\cmidrule(lr){2-3}\cmidrule(lr){4-5}
Property & MAE & RMSE & MAE & RMSE & Gain \\
\midrule
$\omega_{\log}$ (K)              & 24.6   & 55.3   & 22.4   & 51.0   & -8.9\%  \\
$\lambda$                        & 0.0842 & 0.178  & 0.0672 & 0.148  & -20.2\% \\
$E_{\mathrm{form}}$ (eV/atom)    & 0.0461 & 0.0771 & 0.0312 & 0.0536 & -32.3\% \\
$E_{\mathrm{hull}}$ (eV/atom)    & 0.0491 & 0.0825 & 0.0415 & 0.0726 & -15.5\% \\
\midrule
Bulk modulus (GPa)               & 7.54   & 15.7   & 7.26   & 15.5   & -3.8\%  \\
Shear modulus (GPa)              & 9.76   & 20.9   & 8.99   & 21.3   & -7.8\%  \\
Young modulus (GPa)              & 21.0   & 40.4   & 20.4   & 41.5   & -3.0\%  \\
\bottomrule
\end{tabular}
}
\end{table*}

\add{These results show two things. First, the benefit of the pDOS tracks how directly the target is influenced by the electronic structure. It is largest for the electron-phonon coupling constant and for the formation energy, it remains sizeable for the energy above the hull and for $\omega_{\log}$, and it essentially vanishes for the Young modulus. This is the expected behaviour for the superconducting properties. From the McMillan--Hopfield expression for the coupling constant~\cite{McMillan1968Mar, Hopfield1969Oct}, we know how the coupling constant depends on the number of states at the Fermi level, while the vibration component needs to be inferred by the network. As for the elastic response, it is governed by the second derivative of the total energy with respect to strain rather than by the distribution of the states around the Fermi level, and it confirms that the embedding gain does not come simply from giving the network more input features, but from electronic information that is relevant to the target property. Second, and more importantly, a target does not need to be an explicitly spectroscopic quantity for the pDOS to be useful. The formation energy and the distance to the convex hull are thermodynamic quantities, and yet they profit from the embedding as much as, or more than, the superconducting properties. The orbital-resolved projections therefore act as a compact, element- and site-specific descriptor of the local chemistry, which the network exploits whenever the target retains a memory of the electronic structure.}

To validate the generality of the pDOS encoding \add{for small datasets}, we trained \textsc{tuga-sp} on a dataset of $\sim$800 Heusler compounds to predict the nearest-neighbor isotropic exchange interaction $J$ obtained from SIESTA calculations~\cite{Soler2002,Garcia2020} using PBE~\cite{Perdew1996} and post-processed with TB2J~\cite{He2021} using the magnetic force theorem within the LKAG formalism~\cite{Liechtenstein1987}. Heuslers provide a particularly demanding test case because the exchange coupling is determined by subtle $d$--$d$ and $d$--$p$ orbital overlaps that depend sensitively on both the site occupancy and the local electronic environment of the magnetic atoms~\cite{Graf2011}. \remove{This is also a very data-poor case, where the dataset contains $\sim$800 entries.}

Given the small dataset size, hyperparameter optimization was not performed. Instead, the optimal hyperparameters from the \Tc\ study were reused. Without pDOS features, the model achieves a test MAE of 1.24~meV. With pDOS augmentation, this decreases to 1.19~meV, a relative improvement of 4.0\%. This small gain is expected, as for such small datasets, the feature dimensionality of the pDOS vector becomes comparable to the number of training samples, limiting the information the model can reliably extract from the orbital projections. Nevertheless, pDOS-augmented representations still outperform the elemental baseline even under this severe data constraint, confirming that the first-principles electronic fingerprint provides a beneficial inductive bias. We anticipate that larger Heusler datasets or dedicated hyperparameter optimization would yield substantially larger improvements.

\subsection{Attention over orbital projections}
\label{sec:attention}

\begin{figure*}[tbh]
    \centering
    \includegraphics[width=1\textwidth]{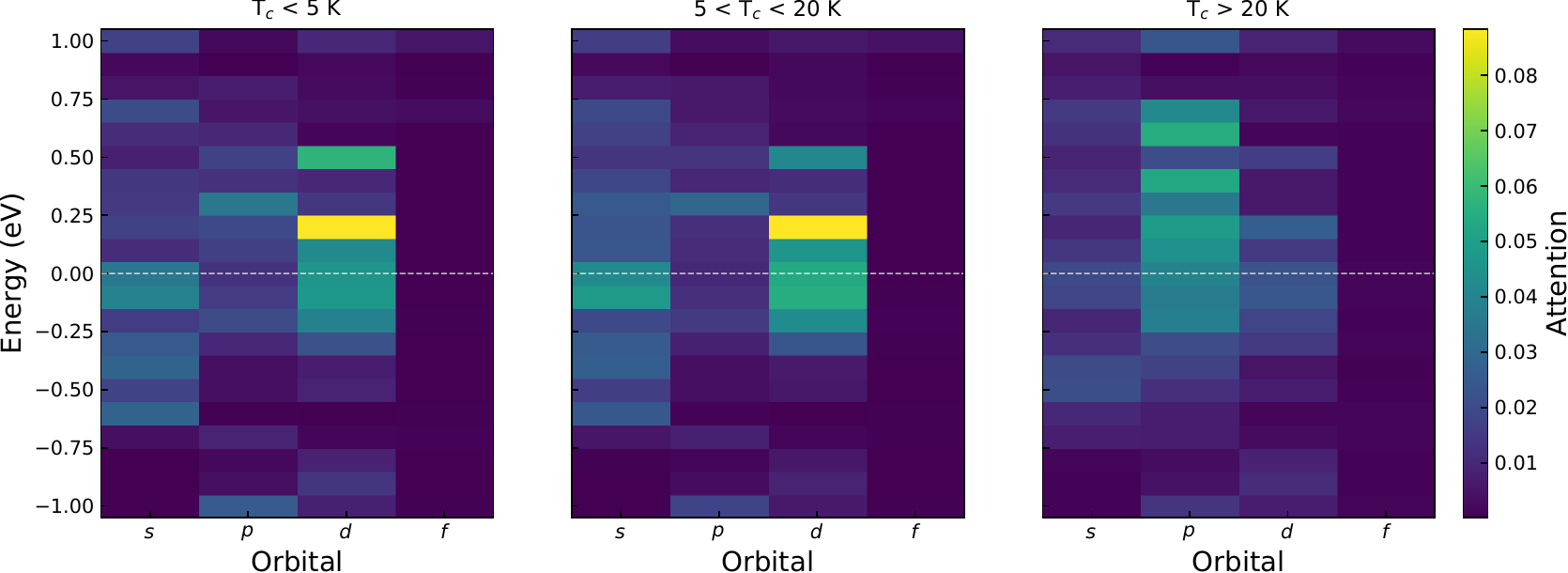}
    \caption{Attention heatmaps averaged over all atomic sites across all $\sim$3\,400 compounds in the superconductivity critical temperature test set, grouped by ranges of \Tc. The vertical axis spans the 21 energy bins from $-1$ to $+1$~eV relative to $E_F$, while the horizontal axis encodes the four orbital channels ($s$, $p$, $d$, $f$). Warmer colors indicate larger learned attention weights.}
    \label{img:attention_all_tc}
\end{figure*}

\begin{figure*}[tbh]
    \centering
    \includegraphics[width=1\textwidth]{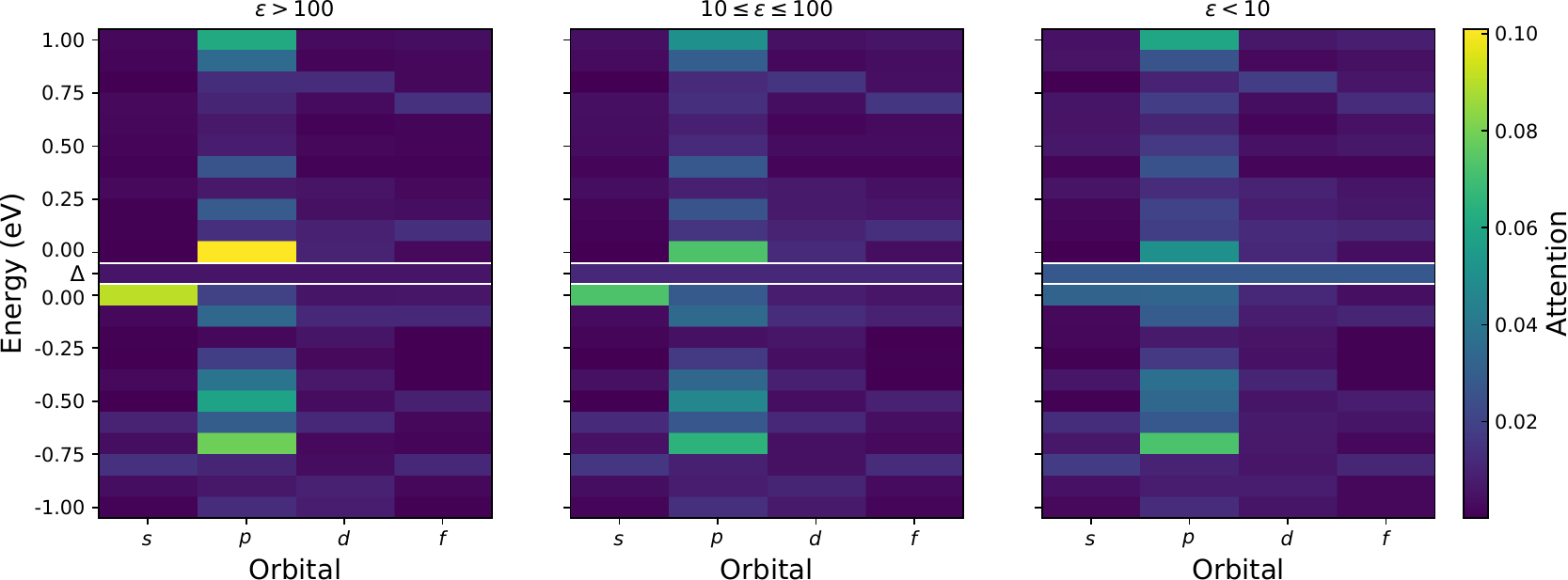}
    \caption{Attention heatmaps averaged over all atomic sites across all $\sim$2\,000 compounds in the dielectric constant test set, grouped by ranges of \epsinf. The vertical axis spans the 11 energy bins from $-1$ to $0$~eV and another 11 energy bins from $0$~eV to $+1$~eV relative to the VBM and CBM respectively, with the gap value being represented in the middle, while the horizontal axis encodes the four orbital channels ($s$, $p$, $d$, $f$). Warmer colors indicate larger learned attention weights.}
    \label{img:attention_all_dec}
\end{figure*}

To gain physical interpretability, we introduce an attention gate mechanism over the pDOS fingerprint. An auxiliary network produces a set of non-negative attention weights $\mathbf{a}_i$ for each atomic site $i$ before multi-layer perceptron (MLP) projection of the pDOS node feature:
\begin{align}
    \mathbf{a}_i &= \mathrm{softmax}\!\left(\mathrm{MLP}_a\!\left(\mathbf{s}_i \,\|\, \mathrm{Pettifor}(Z_i)\right)\right),
    \label{eq:attn_weights}\\
    \tilde{\mathbf{s}}_i &= \mathbf{s}_i \odot \mathbf{a}_i,
    \label{eq:attn_apply}\\
    \mathbf{h}_i &\leftarrow \mathbf{h}_i + \mathrm{MLP}_d\!\left(\tilde{\mathbf{s}}_i\right),
    \label{eq:node_update}
\end{align}
where $\mathbf{s}_i$ is the concatenated pDOS vector (all orbital channels and energy bins), $Z_i$ is the atomic number, $\|$ denotes concatenation, and $\odot$ is the element-wise product. The weights $\mathbf{a}_i$ are conditioned on both the pDOS spectral fingerprint and the chemical identity of the site, through the same Pettifor elemental embedding used; after training this attention weight can be extracted to reveal which orbital channels and energy ranges the model finds most informative for the target property.

The model including this attention mechanism for \Tc\ has an error of 1.01~K, a marginally higher loss compared with the model with no attention. This is an acceptable cost for the physical insight gained. Looking at the global attention maps for \Tc\ (\cref{img:attention_all_tc}), we observe that for \Tc\ values in the lower to mid range the model assigns stronger attention weights near the Fermi level and in the $d$-orbital channel. As we go to the high-\Tc\ regime, the model increases its attention over the $p-$orbitals close to the Fermi level. 

The attention model for \epsinf$^{-1}$ also yields a larger error, with a mean square error of 2.0$\times 10^{-4}$. Looking at the dielectric constant heatmaps (\cref{img:attention_all_dec}) we see that as the \epsinf\ decreases, the model increases the attention to the band gap. This is expected, since the dielectric constant is to some extent inversely related to the band gap~\cite{Marques2011}. As the gap decreases, the domain of possible values for the dielectric constant increases, and the model no longer uses the gap value as a discriminator for the dielectric constant value.

\begin{figure*}[tbh]
    \centering
    \includegraphics[width=0.70\textwidth]{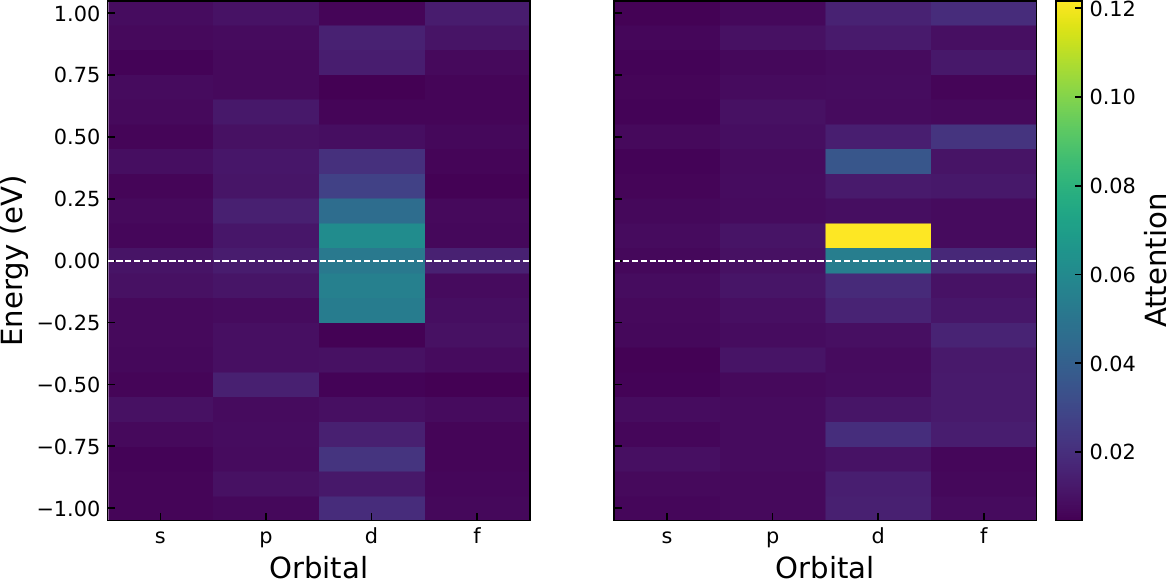}
    \caption{Effect of smearing on learned attention patterns. Attention heat maps averaged over $\sim$400 compounds calculated with a Methfessel-Paxton smearing of 0.2~eV (left) and 0.05~eV (right). Lower smearing produces sharper spectral features and more focused attention peaks; the test MAE changes only marginally (from 1.51~K to 1.50~K), indicating that the model has already extracted the physically relevant information from the broader distributions.}
    \label{img:attention_map_smear}
\end{figure*}

To understand the capability of the model to learn the physics intrinsic to the density of states, we also examined the sensitivity of the learned attention to the DFT smearing parameter (\cref{img:attention_map_smear}). Reducing the Methfessel-Paxton smearing from 0.2~eV to 0.05~eV sharpens the spectral features and produces more concentrated attention peaks, better delineating physically distinct orbital regions. However, the prediction error changes only marginally (1.51~K to 1.50~K), confirming that the model had already retrieved the relevant information even from the broader smeared distributions.

\subsection{Compound-resolved attention maps}
\label{sec:compound_attn}

\begin{figure*}[tbh]
    \centering
    \includegraphics[width=\textwidth]{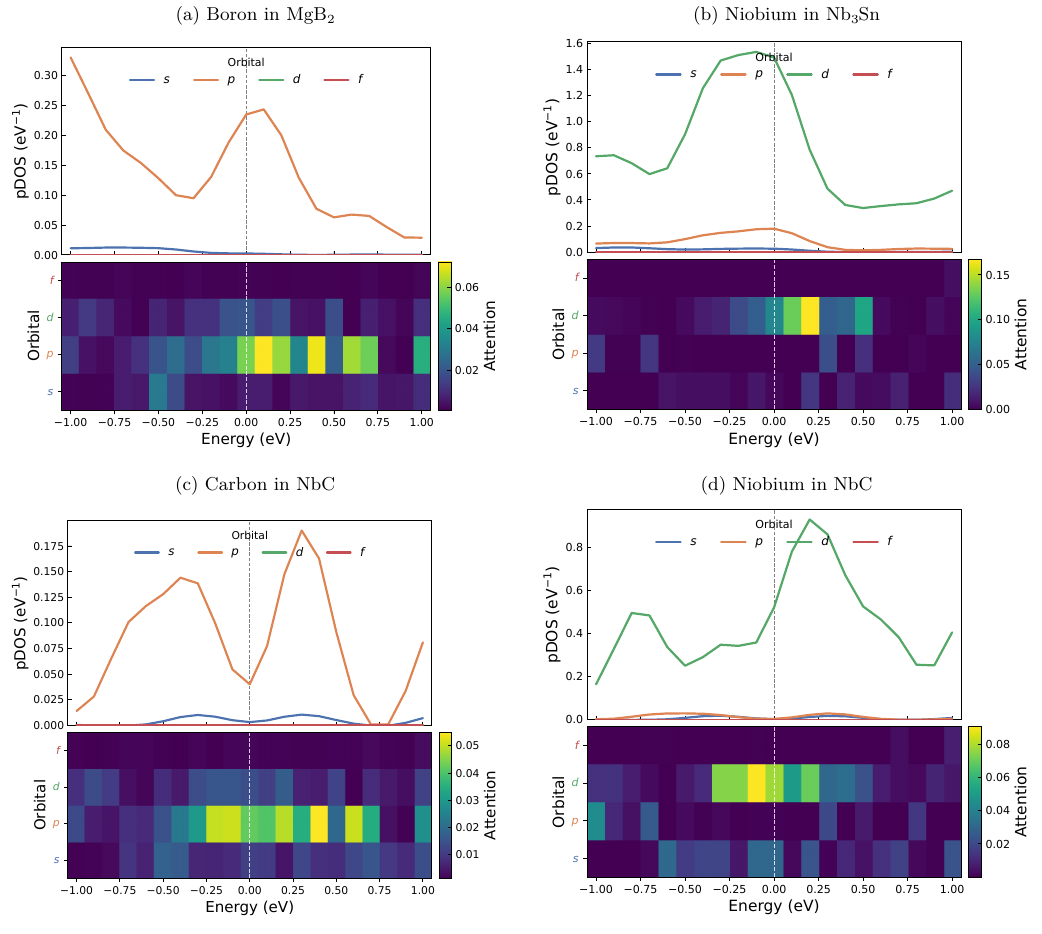}
    \caption{Site-projected density of states and corresponding learned attention weights for known experimental superconductors absent from the training set. a) boron in \ce{MgB2}; b) niobium in \ce{Nb3Sn}; c) carbon in \ce{NbC}; d) niobium in \ce{NbC}. Vertical dashed line marks $E_F = 0$.}
    \label{img:attention_map_species}
\end{figure*}

The use of the attention mechanism allows us to interpret the features for each individual material that the model correlates with the predicted property. We now analyze how the model interprets the electronic structure of specific well-known superconductors (\cref{img:attention_map_species}). 

Superconductivity in \ce{MgB2} (experimental \Tc\ = 39.0~K) originates from $\sigma$-bonding states formed by boron $p_x$ and $p_y$ orbitals near the Fermi level, which couple strongly to the $E_{2g}$ phonon mode~\cite{Nagamatsu2001,An2001,Yu2024}. The model predicts a \Tc\ of 9.5~K, to be compared with the \textit{ab initio} isotropic value of around 17.0~K. In agreement with the theoretical interpretation of superconductivity in this compound, the highest attention weights are assigned to the boron atom, precisely to the $p$-channel in the $[0,  0.4]$~eV window relative to $E_F$ (\cref{img:attention_map_species} (a)). In \ce{Nb3Sn} (experimental \Tc\ = 18.3~K), superconductivity is dominated by the Nb $d$-states near the Fermi energy~\cite{Matthias1954, Labbe1968, Cucciari2025}. Consistently, the model shifts its attention toward higher-energy Nb $d$ conduction states (\cref{img:attention_map_species} (b)), predicting a \Tc\ of 13.7~K.  Finally, in \ce{NbC} rock salt phase (experimental \Tc\ = 11.1~K), superconductivity is also dominated by Nb $4d$-orbitals near $E_F$. However, there is a strong hybridization with C $2p$-states lying slightly below the Fermi level~\cite{Willens1967, Shang2020}. The model predicts a \Tc\ of 15.9~K, paying higher attention in the $d$-states slightly below the Fermi level when compared with \ce{Nb3Sn}. This region overlaps with the same in C $p$ attention heatmap (\cref{img:attention_map_species},  (c) and (d)), proving that the model is both able to generalize its interpretation of the density of states and able to capture the $d$--$p$ hybridization characteristic of the rock-salt niobium carbide.

These examples demonstrate that the model autonomously learns to identify the superconductivity-relevant electronic states in different energy regions depending on the compound. This is an emergent capability arising from learning the mapping between pDOS fingerprints and \Tc\ without any explicit orbital-chemistry supervision.

\section{Discussion}
\label{sec:discussion}

The central finding of this work is that explicit encoding of site-projected orbital densities of states as atomic node features in a crystal GNN yields consistent, substantial improvements in predictive performance across qualitatively different target properties.

Conventional GNN node embeddings are species-level features, where atoms of the same species are encoded identically regardless of whether it sits in a metallic, superconducting, or antiferromagnetic environment. Therefore, conventional GNN is forced to infer the compound-specific electronic environment entirely through iterative message passing over the structural graph. The pDOS embedding bypasses this bottleneck by providing, at the node initialization step, a quantum-mechanically computed spectral fingerprint that already encodes hybridization, charge transfer, oxidation state, and crystal-field effects. Message passing then needs only to aggregate and combine this rich site-level information into a global representation, rather than reconstruct it from structural data alone. The attention-map analysis confirms that the model exploits this information in a physically meaningful way, by selectively attending to the orbital channels and energy windows most relevant to the target properties, discovering, without any type of supervision, the connection between Fermi-level spectral weight and superconductivity, or between gap and band-edge states and dielectric response.

Prior works have attempted to encode structure-wise embeddings in machine-learning models for materials. One class of approaches uses convolutional or vision-type neural networks to encode band-structure images~\cite{Li2025}. However, the re-scaling of band structures to a fixed pixel grid distorts the energy axis for non-cubic systems, introducing artifacts that depend on the arbitrary choice of $k$-path rather than the physical content. A second line of work derives site-specific embeddings from secondary GNNs operating on the local structural environment~\cite{Gupta2026}. While these models avoid static descriptors, they still do not encode the actual electronic state of each site, as that quantum-mechanical information is absent from the geometry alone. The pDOS, by contrast, is a rigorous, site-resolved quantum-mechanical observable computed directly via DFT and adapted to the specific compound. It carries information that cannot be reconstructed from the crystal graph alone. As the accuracy gains achievable by architectural innovation in GNNs have slowed in recent years~\cite{Jain2024}, this input-level enrichment offers a complementary and alternative route to improving performance.

The finding that pDOS features are equivalent to approximately doubling the training data carries a practical consequence of broad importance. In fact, it opens a pathway to train models for properties where labeled data is scarce, such as experimental measurements. Experimental \Tc\ databases and dielectric measurements typically contain a few thousand entries~\cite{Sommer2023, Petousis2017}, exactly the regime where the pDOS features would provide the greatest benefit. The prerequisite is that DFT calculations with pDOS information are available, which is increasingly the case as large high-throughput databases continue to expand~\cite{Cavignac2025}. \add{The choice of such databases, and by consequence the choice of different functionals or methods, does not seem to be determining. Even though different methods might shift/contract the electronic features, they still remain an additional descriptor of such property, as we could see from the results obtained with the different smearings.}


\section{Methods}
\label{sec:methods}

\subsection{TUGA-SP architecture}
\label{sec:tuga}

\textsc{tuga-sp} is an invariant crystal graph neural network that adopts the atom-graph and line-graph representation of ALIGNN~\cite{Choudhary2021}, but replaces its message passing with a transformer-based paradigm. The crystal structure is represented by a primary (atom) graph and its associated line graph. Node features $\mathbf{h}$ represent atoms, edge features $\mathbf{e}$ encode pairwise interatomic distances, and triplet features $\mathbf{t}$ encode bond angles between pairs of edges sharing a common atom.

The atom nodes are initialized with the Pettifor chemical-scale embedding~\cite{Cerqueira2026}, which is then projected by an MLP and normalized. Per-site features are then projected and added to the initial node embeddings.

Interatomic distances are encoded with a Gaussian radial basis function expansion modulated by a cosine cutoff envelope. Bond angles (triplets) are encoded analogously, as Gaussian smeared cosines of the angle.

Each interaction block proceeds in two phases. In the \emph{update phase}, MLP mixers propagate information along the sequence $\mathbf{h} \to \mathbf{e} \to \mathbf{t}$, with skip connections carrying the pre-update features: each edge is updated from its two endpoint nodes, and each triplet from its two constituent edges. In the \emph{message-passing phase}, a graph-transformer convolution aggregates information in the reverse direction $\mathbf{t} \to \mathbf{e} \to \mathbf{h}$, so that updated triplet features mediate edge aggregation (line graph) and updated edge features mediate node aggregation (atom graph). In each convolution, the mediating feature enters both the attention weights and the aggregated messages. Multiple interaction blocks are stacked, the number being selected by hyperparameter optimization.

After the final interaction block, attention-weighted pooling is applied independently to each set of graph elements, yielding pooled node, edge, and triplet representations that are concatenated into a global crystal representation $[\mathbf{h} \,\|\, \mathbf{e} \,\|\, \mathbf{t}]$. A final MLP head maps this representation to the target property.

\subsection{Data}
\label{sec:data}

The structure and thermodynamic data used in the training of the models was taken from the Alexandria database~\cite{Cavignac2025}.

\add{All DFT calculations were performed using the Vienna Ab initio Simulation Package (VASP)~\cite{Kresse1996, Kresse1996_2} with the Perdew-Burke-Ernzerhof (PBE) exchange-correlation functional~\cite{Perdew1996} and Projector augmented-wave (PAW) pseudopotentials~\cite{Blochl1994, Kresse1999}. The dataset for the critical temperature was computed with the materials project parameters. The energy cutoff used was 520~eV. The Brillouin zone was sampled using a $\Gamma$-centered k-point grid with a density of 1000 k-points per reciprocal atom. A Gaussian smearing was applied with a smearing width of $\sigma$= 0.2~eV. The calculations were converged to forces smaller than 0.005 eV ~\AA$^{-1}$. 
The dataset for the remaining properties (including the critical temperature with the 0.05~eV smearing) was computed following the MATPES settings~\cite{Kaplan2025}. The energy cutoff used was 680~eV. The Brillouin zone was sampled using a $\Gamma$-centered k-point grid with a spacing of 0.22~\AA$^{-1}$. A Gaussian smearing was applied with a smearing width of $\sigma$= 0.05~eV. The calculations were performed with spin polarization, using convergence criteria of $10^{-7}$~eV for the total energy and $5$~meV~\AA$^{-1}$ for the forces. The smearing correction introduced in MATPES was not applied.
The site- and orbital-resolved densities of states were obtained by projecting the Kohn--Sham orbitals onto spherical harmonics ($lm$ channels) centred on each atom, using the PAW projector functions within the augmentation spheres (\texttt{LORBIT}~=~11). As this projection neglects the interstitial region, the summed partial densities of states do not exactly recover the total density of states. The thermodynamic properties considered are present in the Alexandria database, computed with the mentioned settings.

\subsection{Superconducting critical temperature}

Density-functional perturbation theory (DFPT) calculations were performed with \textsc{Quantum ESPRESSO} (versions 6.8 and 7.1)~\cite{Giannozzi2009,Giannozzi2017} using the PBEsol exchange--correlation functional~\cite{Perdew2008}. Structural relaxations employed $\Gamma$-centered $k$-point meshes corresponding to 3000 $k$-points per reciprocal atom, with odd grid dimensions increased to the next even integer. Convergence thresholds of $10^{-8}$~a.u. for total energies, $10^{-6}$~a.u. for forces, and $5\times10^{-2}$~kbar for stresses were adopted.
Electron--phonon coupling calculations were carried out using the double-grid technique, where the coarse electronic mesh was obtained by doubling the relaxation $k$-point grid in each reciprocal direction and the fine mesh by quadrupling it. Phonons were sampled on $\mathbf{q}$-point meshes equal to one half of the relaxation $k$-point grid in each direction. The Eliashberg function $\alpha^2F(\omega)$ was computed using double-$\delta$ integration with a Methfessel--Paxton broadening of 0.03~Ry. The critical temperature value was computed with the Allen-Dynes equation, with a coulomb interaction of $\mu^* = 0.1$, which is the typical criteria for theoretical \Tc{} values.}

\subsection{Static dielectric constant}
\add{The optical dielectric constant $\epsilon_\infty$ was obtained from density-functional perturbation theory as implemented in VASP~\cite{Kresse1996, Kresse1996_2} (\texttt{LEPSILON = .TRUE.}), which yields the electronic contribution to the static macroscopic dielectric tensor, including local-field effects at the DFT level. The target value is the average of the diagonal elements of this tensor, $\epsilon_\infty = \mathrm{Tr}(\boldsymbol{\epsilon}_\infty)/3$.}

\subsection{Isotropic exchange in Heusler compounds}

\add{The nearest-neighbor isotropic exchange interactions $J_{ij}$ were obtained with the TB2J package~\cite{He2021}, which applies the magnetic force theorem (in the LKAG formalism) to the Kohn-Sham Hamiltonian expressed in a local orbital basis. The underlying first-principles calculations were performed within DFT using the PBE functional~\cite{Perdew1996}, as implemented in SIESTA (v 5.4.1)~\cite{Soler2002, Garcia2020}, with a double-$\zeta$ polarized (DZP) basis set for all atoms. The converged Kohn-Sham Hamiltonian and overlap matrices were then passed to TB2J. The extracted couplings define a Heisenberg Hamiltonian,
\begin{equation}
H = - \sum_{i \neq j} J_{ij}\,\mathbf{S}_i \cdot \mathbf{S}_j,
\end{equation}
where $\mathbf{S}_i$ is a normalized spin vector ($|\mathbf{S}_i|=1$). Because the magnetic force theorem evaluates the exchange parameters at the unit-cell level, no supercells or additional self-consistent calculations for different magnetic configurations are required.}

\add{All remaining properties correspond to the benchmark dataset used by MatCalc~\cite{MatCalc}.}

\section*{Data availability}

\add{The crystal structures, total and formation energies, energies above the convex hull, and site- and orbital-projected densities of states used in this work are openly available from the Alexandria database (\url{https://alexandria.icams.rub.de})~\cite{Cavignac2025}. The elastic moduli were taken from the publicly available benchmark dataset of MatCalc~\cite{MatCalc}.} 

\section*{Code availability}

\add{The source code of \textsc{tuga-sp} used in this work is available at \url{https://github.com/ppdebreuck/tuga-sp} 
The DFT calculations were performed with VASP, which is available under licence, \textsc{Quantum ESPRESSO} 6.8 and 7.1, and SIESTA 5.4.1, and the exchange constants were extracted with TB2J. Hyperparameter optimization was performed with Optuna. All parameters used to generate and analyze the data are given in the Methods section.}

\section*{Acknowledgements}

The authors gratefully acknowledge the computing time made available to them on the high-performance computers Noctua and Otus at the NHR Center Paderborn Center for Parallel Computing (PC2). The authors gratefully acknowledge the scientific support and HPC resources provided by the Erlangen National High Performance Computing Center (NHR@FAU) of the Friedrich-Alexander-Universität Erlangen-Nürnberg (FAU) under the NHR project k114eb. Part of the calculations were performed on the HPC cluster Elysium of the Ruhr University Bochum, subsidised by the DFG (INST 213/1055-1).

\section*{Funding}

P.R.P., M.F., and M.A.L.M. were supported by a collaboration between the Kavli Foundation, Klaus Tschira Stiftung, and Kevin Wells, as part of the SuperC collaboration, and by the Simons Foundation through the Collaboration on New Frontiers in Superconductivity (Grant No. SFI-MPS-NFS-00006741-10).

\section*{Competing interests}

The authors declare no competing financial or non-financial interests.

\renewcommand{\bibsection}{\section*{References}}
\bibliography{bib.bib}

\end{document}